\documentclass[12pt,epsfig]{article}

\usepackage{amssymb,epsfig}

\usepackage{psfrag}

\psfrag{v0}{{$\vec 0$}}
\psfrag{Bup}{$((1- \zeta)\,n^+, \vec B /(1-\zeta) )$}
\psfrag{Bdown}{$(\zeta \;n^+, -\vec B /\zeta )$}
\psfrag{2B}{$2\vec B$}
\psfrag{Bd}{$\vec B$}
\psfrag{z}{$z$}
\psfrag{bz}{$1-z$}
\psfrag{-2B}{$-2\vec B$}
\psfrag{{ReF}}{Re $F$}
\psfrag{{Im F}}{Im $F$}
\psfrag{b}{$|\vec B|$}

\newcommand{\be}{\begin{equation}}

\newcommand{\ee}{\end{equation}}

\newcommand{\bea}{\begin{eqnarray}}

\newcommand{\eea}{\end{eqnarray}}
\newcommand{\bv}{\vec{b}}

\newcommand{\p}{\vec{p}}

\newcommand{\D}{\vec{\cal D}}
\newcommand{\B}{\vec B}
\newcommand{\dB}{\delta \vec B}

\begin{document}

\begin{center}

{\LARGE \bf   Impact Representation of Generalized Distribution Amplitudes}
\vspace{1cm}

{\sc B.~Pire$^a$ and L. Szymanowski$^b$,}
\\[0.5cm]
\vspace*{0.1cm}
{\it   $^a$CPhT, {\'E}cole Polytechnique, F-91128 Palaiseau,
France\footnote{
  Unit{\'e} mixte C7644 du CNRS.}
                       }
\\
$\mbox{}$\\
 {\it
 $^b$So{\l}tan Institute for Nuclear Studies,
Ho\.za 69,\\ 00-681 Warsaw, Poland
                       } \\[1.0cm]

\vskip2cm
\end{center}

\begin{abstract}
{\small We develop an impact representation for the generalized
distribution amplitude which describes the exclusive hadronization of a
quark-antiquark pair to a pair of mesons. Experiments such as
$\gamma^* \gamma \to \pi \pi $ and $\gamma^* N \to \pi \pi N'$ are
shown to probe the transverse size of the hadronization region  of the
quark antiquark pair that one can interpret as the transverse overlap of the
two emerging
mesons. An
astonishing feature
of this description is that low energy $\pi \pi$ phase shift analysis can
be used for understanding some properties of quark hadronization process.
}
\end{abstract}

\vskip1cm

\noindent
{\bf 1.} A considerable amount of theoretical and experimental work is
currently being devoted to the study of generalized parton
distributions \cite{GPD,rev}, which are defined as Fourier transforms of
matrix elements between different hadron states, such as:
$$
\int d\lambda e^{i\lambda x(p+p').n} \langle N'(p',s') |\bar\psi(-\lambda
n)(\gamma \cdot n)
\psi(\lambda n) | N(p,s) \rangle
$$
and their crossed version which describe the exclusive hadronization of
a
$q
\bar q$ or $g g$ pair in a pair of hadrons, a pair of $\pi $ mesons for
instance. These generalized distribution amplitudes (GDA) \cite{DGPT},
defined in the quark-antiquark case, as
\be
\Phi_q^{\pi\pi} (z,\zeta,W) =\int\frac{d\lambda}{2\pi} e^{-i\lambda
z(p+p')n^*}
\langle \pi(p') \pi(p) |\bar\psi(\lambda n^*)(\gamma \cdot n^*)
\psi(0) | 0) \rangle
\label{def}
\ee
where $W$ is the squared energy of the $\pi \pi$ system,
$W^2=(p+p')^2$,
are  non perturbative parts entering in a factorized way \cite{AFm} the
amplitudes
 of the light cone
dominated processes
$$
\gamma^* \gamma \to \pi \pi
$$
which may be measured \cite{DGP} in electron positron colliders of high
luminosity and
$$
\gamma^* N \to \pi \pi N'
$$
which is under intense study at HERA and JLab \cite{HERAJLAB} and has been
the subject of
some recent
theoretical works \cite{Polyakov:1998ze, Hagler}.
This new QCD object allows to treat in a consistent way
the final state interactions of the meson pair. Its phase is related to the
 phase of $\pi \pi $ scattering amplitude, and thus contains information on
the resonances which may decay in this channel.

The measurements of GPDs and GDAs are expected to yield important
contributions to our understanding of how quarks and gluons assemble
themselves into hadrons.

A peculiar feature of GPDs has recenty been the subject of intense
investigation \cite{femto,MD}. Apart from longitudinal momentum fraction
variables, GPDs also depend on the momentum transfer $t$ between the initial
and final hadrons. Fourier transforming this transverse momentum
information leads to information on the transverse location of quarks and
gluons in the hadron. Real-space images of the target can thus be obtained.
  Spatial resolution is determined by the virtuality
of the incoming photon.

In this note, we show that a similar argument can be developed
for the $W^2$ dependence of the GDAs. By developing essentialy the same
techniques as in Ref. \cite{MD} we  Fourier
transform the $2-$dimensional transverse momentum behavior of the GDAs into a
transverse impact coordinate $\vec b$ representation. We
then give the physical interpretation of the $\vec b$ behaviour of this new
function. We show that general theorems allow us to get some
insight on this $\vec b$ behaviour; we discuss the consequences of such
representation in a simple analytical model and briefly comment on future
phenomenology.


Throughout this paper we restrict our study to the quark-antiquark
initial state and to the two
pion case, but gluon contributions may be studied along similar lines.
The remarks on possible other channels we present in the discussion.
Also we do not include the QCD evolution
of the GDAs, which doesn't influence their dependence on $W$,
hence the $\vec b$-dependence of their Fourier transform.

\vskip.1in
\noindent
{\bf 2.} Let us describe the kinematics of
the $q\,\bar q \to \pi\,\pi$
 process under consideration.
We parametrize all momenta by means of the Sudakov decomposition with two
light-like
vectors
$n$ and $n^*$ satisfying the conditions $n \cdot n^* = 1$.
The parametrization of pions momenta $p$ and $p'$ reads
\bea
\label{kin}
\nonumber \\
&&p=\zeta\;n + \frac{\p^{\;2} + m_\pi^2}{2\zeta \,}n^* + p_\perp
\nonumber \\
&&p'= \bar \zeta \;n + \frac{\p^{\;'\,2} + m_\pi^2}{2 \bar \zeta } n^* +
p'_\perp
\nonumber \\
&&p_\perp^2= -\p^{\;2}\;,\;\;\;\;\;\bar \zeta = 1-\zeta\;,
\eea
where $m_\pi$ is the pion mass.
\noindent
The invariant mass squared of the two pion system equals then
\be
\label{invmass}
W^2= \frac{m_\pi^2}{\zeta \bar \zeta} + \zeta \bar \zeta\; \left(
 \frac{\p}{\zeta}- \frac{\p^{\;'}}{\bar \zeta}  \right)^2\;,
\ee
i.e. its dependence on transverse momenta of pions enters via the
modulus squared of the two dimensional vector
 $\D$
\be
\label{D}
\D=  \frac{ \p}{\zeta} -\frac{\p^{\;'}}{\bar \zeta}\;.
\ee
The crucial property of this vector is its invariance under transverse
boosts \cite{Kogut:1969xa}
\bea
\label{tboost}
&& \zeta \to \zeta \;,\;\;\;\;\;\;\;\p \to \p - \zeta\,\vec
v\;,\;\;\;\;\;\;\;
\p^{\,'} \to \p^{\,'} - \bar \zeta\,\vec v \;,
\nonumber \\
&&\D \to \D \;.
\eea
The definition of such a vector $\D$ is of course not  unique: all
vectors which
differ >from the one given by Eq. (\ref{D}) by a function of $\zeta$ are
equally good.
Nevertheless, the physical conclusions which we draw below do not depend on
this arbitrariness.
\vskip.1in
\noindent
{\bf 3.} The description of hadrons in coordinate space
is done through the use of  wave
packets with a finite width to avoid the appearence of infinities in the
intermediate stages. We define first these wave packets in momentum space
as
\be
\label{wavep}
|p^+ , \p \rangle_\Phi =
\Phi(p^+,\p)|p \rangle\;,
\ee
 where $\Phi(p^+,\p)$ is a function describing the shape of the wave packet. In
analogy with Ref. \cite{MD} we use a simple Gaussian ansatz for this shape
function, which permits to perform the calculations analytically. We define
\bea
\label{Phi}
&&\Phi(p)= p^+\delta(p^+ - p^+_0)\Phi_\perp(\p) \;,
\nonumber \\
&&\Phi_\perp(\p) = G(\p, \frac{1}{\sigma^2})\;,
\eea
where
\be
\label{Gauss}
G(\vec x, \sigma^2) = e^{-\frac{\vec x^{\;2}}{2 \sigma^2}}\;,
\ee
$\sigma$ being the transverse width of the shape function. We can now
Fourier-transform in the $2-$dimensional transverse space.
Thus we define the wave packet as
\be
\label{wavefinal}
|p^+ , \bv_0\rangle_\sigma = \int \frac{d^2\p}{16\pi^3}
e^{-i\p\bv_0}\;G(\p, \frac{1}{\sigma^2})|p \rangle \;,
\ee
which describes a state with a definite light-cone momentum fraction, but
localized around
the transverse position
$\bv_0$ up to the width
$\sigma$;  its normalization is given by
\be
\label{norm}
_\sigma\langle p^{'+}, \bv^{\;'}| p^+, \bv \rangle_\sigma =\frac{1}{16
\pi^2\sigma^2}\;G(\bv-\bv^{\;'},\sigma^2) \stackrel{\sigma \to 0
}{\longrightarrow}
\frac{1}{4\pi}p^+ \delta(p^+ - p^{'+})\delta^2(\bv
-
\bv^{\;'})\;.
\ee
%
\begin{figure}[t]
\centerline{\epsfxsize7.0cm\epsffile{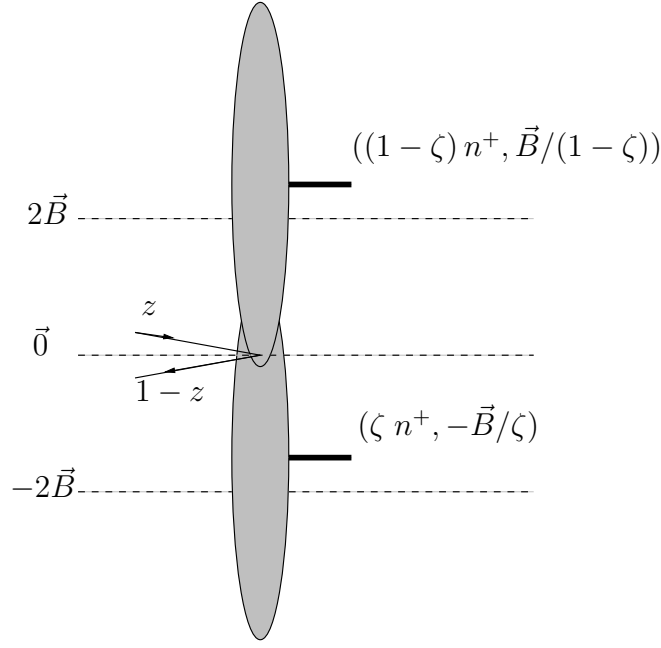}}
\caption[]{\small
Representation of GDA in the impact parameter space for $\zeta \geq 1/2$.
>The pions are characterized by large light-cone components and  transverse
 impact representation coordinates.
 }
\label{fig:1}
\end{figure}
%
\vskip.1in
\noindent
{\bf 4.} Let us now define the 2-dimensional Fourier transform of the
GDAs which will allow us to interpret experimental data in a new way. For
the operator defining the GDA in Eq. (\ref{def}), we get

\be
\label{O}
\hat 0(z, \vec r) =\int\frac{d\lambda}{2\pi} e^{-i\lambda
z(p+p')n^*}
\bar\psi(\lambda n^*, \vec r)(\gamma \cdot n^*)
\psi(0, \vec r)\;,
\ee
which allows to define the GDA for the wave packets defined above as:
\bea
\label{mel}
&&\tilde\Phi_ \sigma^{\pi \pi} (z,\zeta,\bv,\bv^{\;'}) =
_\sigma\langle  \pi(p^{'+},
\bv^{\;'}\,)  \pi(p^{+}, \bv) |
\hat O(z, \vec 0)|0\rangle_\sigma
\nonumber \\
&&=
\int \frac{d^2\p}{16\pi^3}
\frac{d^2\p^{\;'}}{16\pi^3}\;e^{i\p^{\;'}\bv^{\;'} +i\p \bv}
G(\p^{\;'},\frac{1}{\sigma^2}) G(\p,\frac{1}{\sigma^2}) \Phi^{\pi
\pi}(z,\zeta,W(\D))\;.
\eea

\noindent A simple
change of
variables and straightforward gaussian integral lead to the result

\bea
\label{result}
&&
\tilde\Phi_ \sigma^{\pi \pi} (z,\zeta,\bv,\bv^{\;'})  = \int \frac{d^2
\D}{16\pi^3}\;\frac{\zeta^2\,\bar \zeta^2}{8\pi^2 \sigma^2 (\zeta^2 +\bar
\zeta^2)}
\nonumber \\
&& e^{-i\B \D +i\D \dB\,\frac{\bar \zeta^2}{\bar \zeta^2 + \zeta^2}}
\;G(\D, \frac{\zeta^2 +\bar \zeta^2}{\sigma^2 \zeta^2 \bar \zeta^2})\;
G(\dB, \sigma^2(\bar \zeta^2 +\zeta^2))\;
\Phi^{\pi \pi}(z,
\zeta,W(\D))\;,
 \eea
where the transverse positions of the two mesons are given by
\be
\label{B}
\bv = {\frac{1}{\zeta}}\left(\dB -\B  \right)\;,\;\;\;\;\;\;
\bv^{\;'}={\frac{1}{\bar \zeta}}\,\B\;.
\ee
The first observation coming from Eqs. (\ref{result}, \ref{B}) is that,
analogously as in the GPD case
studied in \cite{MD}, the transverse positions of the two produced mesons
($\bv$, $\bv^{\;'}$) depend on $\zeta$.
In particular,  in the limit of vanishing width
$\sigma$,  we have
\be
\label{sigma0}
\tilde\Phi_ \sigma^{\pi \pi}(z, \zeta, B)
\stackrel{{\sigma \to 0}}{=} \frac{\zeta^2 \,\bar
\zeta^2}{4\pi}\delta^2(\dB) \int
\frac{d^2\D}{16\pi^3}\;e^{-i\B\D}\;\Phi^{\pi \pi}(z,\zeta,W(\D))\;.
\ee
Rewriting Eq. (\ref{result}) with vanishing  smearing function, $\dB =0$,
we obtain
\bea
\label{dB0B}
&& \int \frac{d^2 \D}{(2\pi)^2}\;
e^{-i\B \D}\;G(\D,{\cal D}^{\;2}_\sigma)\,\Phi^{\pi\pi}(z,\zeta,
W(\D))
\nonumber \\
&&= \frac{32\pi^3\sigma^2(\bar \zeta^2 +\zeta^2)}{\zeta^2\,\bar \zeta^2}
  \mbox{}_\sigma\langle  \pi(p^{'+},{\frac{\B}{ \bar \zeta}}) \pi(p^{+},
- {\frac{\B}{\zeta}} ) |
\hat O(z, \vec 0)|0 \rangle_\sigma
\eea
where
\be
\label{Ds}
{\cal D}_\sigma^2 = \frac{\zeta^2 +\bar \zeta^2}{\sigma^2\,\zeta^2\,\bar
\zeta^2}\;.
\ee
The result given
by Eq. (\ref{dB0B}) is the GDA analog of  relations appearing in studies
of GPDs, see Eqs.~(16, 17) in Ref.~\cite{MD}.
One can perform the angular integration in
Eq. (\ref{dB0B}) since the Gauss function and the GDA only depend on
$\D^2$. One
gets
\be
\label{Bessel}
\int \frac{d\,\D^2}{4\pi}J_0(|\B||\D|) \;G(\D,{\cal
D}^{\;2}_\sigma)\,\Phi^{\pi\pi}(z,\zeta,
W(\D)) = \frac{32\pi^3\sigma^2(\bar \zeta^2 +\zeta^2)}{\zeta^2\,\bar \zeta^2}
\tilde \Phi^{\pi \pi}_\sigma(z,\zeta, \bv, \bv^{\;'})\;,
\ee
where $J_0$ is the Bessel function, and
\be
\label{shiftb}
\bv=-{\frac{\B}{\zeta}}\;,\;\;\;\;\;\;
\bv^{\;'}={\frac{\B}{1- \zeta}}\;.
\ee

\vskip.1in
\noindent
Let us also note now that due to the normalization of wave packets
(\ref{norm}),
$\sim 1/\sigma^2$,
the right hand side of Eq. (\ref{Bessel}) has a finite
limit for vanishing width $\sigma$.

\vskip.1in
\noindent
{\bf 5.} The physical
picture which emerges from Eq. (\ref{dB0B}) is presented in Fig. 1.
The operator $\hat O(z,\vec 0)$ probes the partons at the transverse
position $\vec 0$ in the two pion system. Since the natural
partonic description of this process is in the infinite momentum frame, the two
emerging pions will be quickly separated in longitudinal space, at least
for a non
symmetric configuration ($\zeta
\neq \bar \zeta $) and we may consider that their transverse separation is
frozen just after production. This allows to identify the transverse
locations of
outgoing mesons as the transverse coordinates derived above.
In order to be
more
specific,  let us  visualize the information in  a simplistic
quark-antiquark toy  model
for the pion structure. A $u \bar u$ initial state (produced by the
$\gamma^* \gamma$ collision for instance)  begins to hadronize at impact
parameter $\vec 0$. The QCD vacuum then produces
a  $d \bar d$ pair in
order to allow
color rearrangement and the formation of the two final pions.
This production process proceeds in
a non-local way, i.e. the transverse positions  of the emerging
$d$ and  $\bar d$ quarks are not at the same point.
The final state pions are then produced at transverse
localizations around
$2\B$ and $-2\B$ (for almost  symmetric configuration of the produced pions
{\it i.e.}
with $\zeta \sim 1/2$).
 If $\zeta$ increases, which is the case shown in
Fig.~1, the
transverse localizations of pions change by  amounts proportional to
the difference $(\zeta -1/2)$, namely $2(\zeta -1/2)/(1-\zeta)$ and
$2(\zeta -1/2)/\zeta$,
respectively, directed towards the top of the figure.
 If $\zeta$ decreases the
shifts of the tranverse positions of pions occur in the opposite direction.
Note however that the values of these shifts are different for both
produced pions.  One should note also that this behaviour doesn't depend
on the longitudinal variable $z$ of quarks entering the operator $\hat O$,
therefore it
persists in the full scattering amplitude, obtained after convolution in
$z$ of the l.h.s.
of Eq. (\ref{dB0B}) with the perturbative hard part of a process.


The $\vec b-$dependence measures the transverse size of the hadronization
 process.
 If the transverse distance $D$ between the pions transverse positions
$\vec b$ and $\vec b^{\;'}$ ( the interaction point being at position
$\vec
0$)
  is larger then the typical pion radius then
 the GDA of the two pion system describes a highly non-local (in
transverse
 space) phenomenon which depends much on the long distance structure of
the
 QCD vacuum. If this transverse distance $D$ is very small (which
corresponds to a large invariant mass of two pion system), a perturbative
 treatment can be applied and the two meson distribution amplitude can be
 expressed in terms of the distribution amplitudes (DA)  of each separate
 pion. In the case when the distance $D$ is smaller than the typical pion
 radius, but still large enough for non-perturbative physics to dominate,
we
 have to deal with a complicated two pion system (at least in the
infinite
 momentum frame) without reference to the DAs of each separate pion. The
 impact representation of GDAs serves as a tool for studying consistently
these three regimes.

\vskip.1in
\noindent
{\bf 6.} The phenomenology of this impact representation is quite rich.
Without developing it in this short note, let us point out a very peculiar
feature, namely the fact that {\it phase shift analysis at low energy can lead
to the understanding of quark hadronization
}, although the concept of quarks is not suitable for the
description of
$\pi \pi $ elastic scattering in this energy domain. This  astonishing
conclusion comes from the following argument. The Watson theorem allows to
derive
the low $W^2$ behavior of the phase of the GDA \cite{Polyakov:1998ze}.
Indeed this theorem
based on the requirement of the unitarity of the $S-$matrix implies that
the phase of the GDA
equals the phase of the
$\pi \pi $ scattering matrix, provided there is no inelastic channel open.
The  phase of $\pi
\pi$ scattering has been the subject of intense study in the last 50 years
(for instance see
 Ref.~\cite{pipi}) and quite a consensus has been achieved at least at energies
below
1~GeV. Through a dispersion relation of the Omn\`es Muskhelishvili type,
one is thus able to
derive a plausible model \cite{DGP,Hagler} of the $W^2$ dependence of the
GDA.  For
instance in the P-wave channel, a single Breit Wigner representation with a
$\rho$ meson pole
is quite reasonable for the low $W^2$ region. If we assume that this region
gives the dominant
contribution to the Fourier integral defining $\tilde \Phi^{\pi \pi}
(z,\zeta,B)$, one
understands that the $\pi \pi$ scattering data collected at low energies
convey information on the transverse
picture of the hadronization process $q \bar q \to \pi\,\pi$.
 This information is not
complete since the
$W^2$ behaviour of $\Phi^{\pi \pi}$ in the medium energy region cannot be
derived from $\pi
\pi$ scattering data. However, standard arguments on the Fourier transformation
allow to trust the transverse location information derived from data with
$W^2$ up to
1~GeV$^2$ as a reliable picture with a resolution of the order of
1~GeV$^{-1}$ {\it
i.e.} one fifth of a femtometer. Finally, the large
$W^2$ region is of very difficult experimental access, but only leads to
further
refinement of the knowledge of the impact picture. One may however trust a
perturbative
estimate \cite{DiehlKroll} of the GDA in this domain and implement it in the
Fourier
transform.


\begin{figure}[t]
\begin{minipage}[t]{72mm}
\centerline{\includegraphics[scale=0.95]{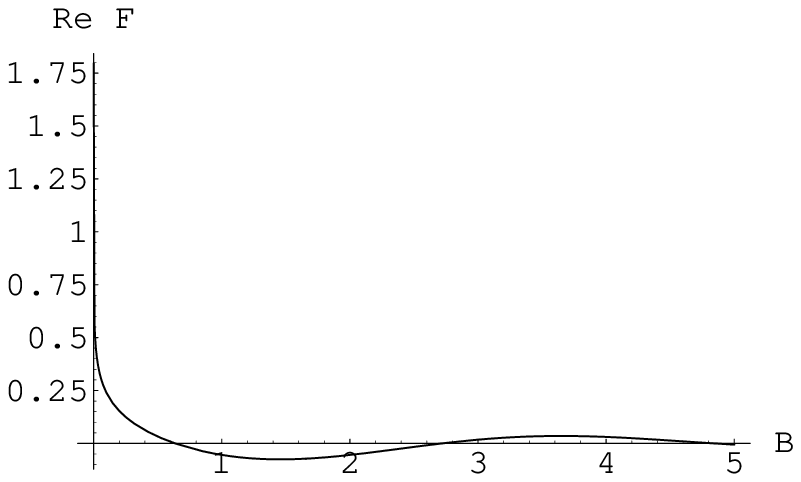}}
\end{minipage}
\hspace{\fill}
\begin{minipage}[t]{72mm}
\centerline{\includegraphics[scale=0.95]{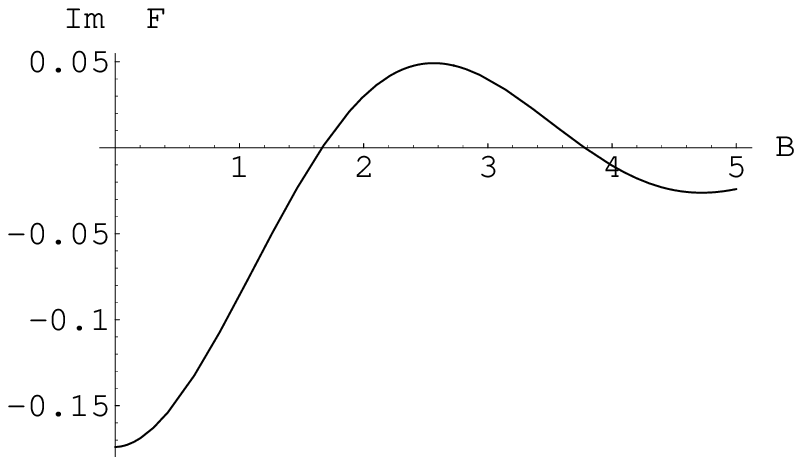}}
\end{minipage}
\caption{{\protect\small Impact parameter dependence of the function $F$
defined
 in Eq. (\ref{F}) for $\zeta = 0.4$ and $z=0.5$ with the Breit-Wigner ansatz.
$|\vec B|$ units are GeV$^{-1}$.
 }}
\label{04}
\end{figure}



\begin{figure}[t]
\begin{minipage}[t]{72mm}
\centerline{\includegraphics[scale=0.95]{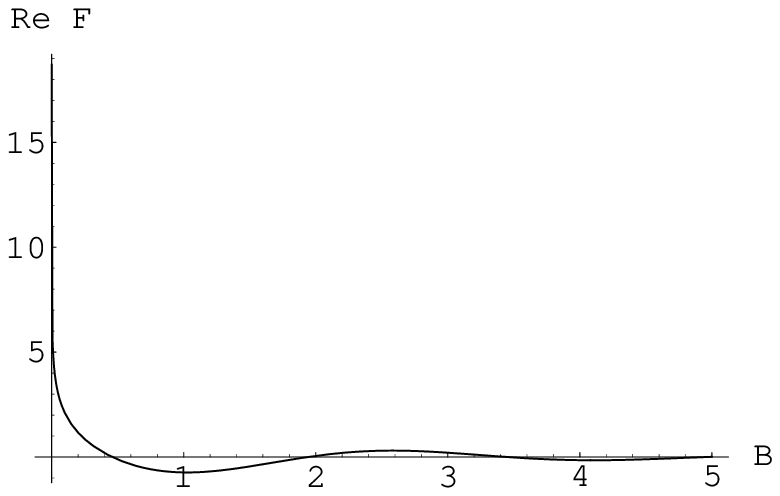}}
\end{minipage}
\hspace{\fill}
\begin{minipage}[t]{72mm}
\centerline{\includegraphics[scale=0.95]{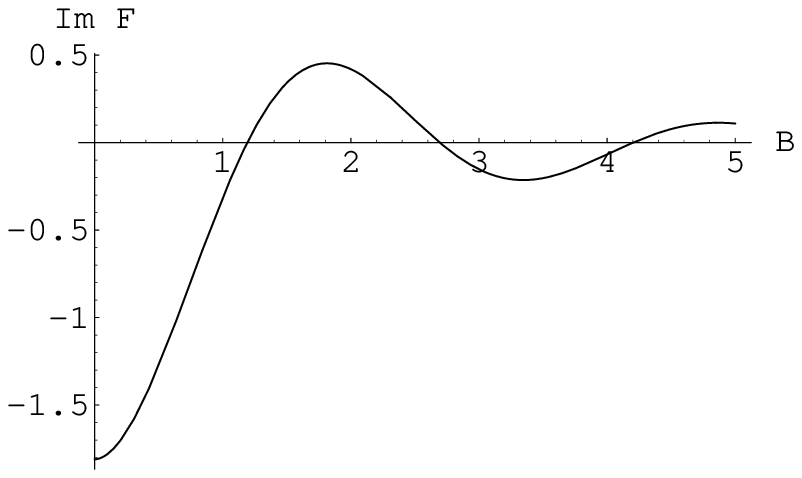}}
\end{minipage}
\caption{{\protect\small The same as in Fig.~\ref{04} for $\zeta=0.1$.
 }}
\label{01}
\end{figure}


\vskip.1in
\noindent
{\bf 7.} As an example we consider the simple model with the outgoing pions in
the $p-$wave, being
described as a $\rho$ meson. The $W$-dependence of the GDA enters in this case
via  a simple Breit-Wigner
form
\bea
&&\Phi^{\pi\pi}(z,\zeta,W)=6z\bar z (2\zeta - 1 ) F_\pi(W)
\nonumber \\
&&F_\pi(W)= \frac{m_\rho^2}{m_\rho^2-W^2-i\Gamma_\rho\,m_\rho}\;,
\eea
with $m_\rho=0.773\;$GeV, $\Gamma_\rho=0.145\;$GeV. We define the function $F$
as
\bea
\label{F}
&&F(z,\zeta,|\vec B|) = \int\limits_0^\infty \frac{d\,\D^{\;2}}{4\pi}\,
\;J_0(|\vec B||\D|)\;\Phi^{\pi\pi}(z,\zeta,W)
\nonumber \\
&& = -\frac{3z\bar z\,(2\zeta - 1)m_\rho^2}{\pi\,\zeta \bar
\zeta}\;K_0\left(|\vec
B|\sqrt{\frac{1}{\zeta\,\bar \zeta} \left(\frac{m_\pi^2}{\zeta \bar \zeta}
-m_\rho^2
+ i\Gamma_\rho\,m_\rho \right)}\right)\;,
\eea
which corresponds to Eq.~(\ref{dB0B}) in the limit of vanishing $\sigma$.
On Figs. \ref{04} and  \ref{01} we plot
the real and imaginary parts of the function $F$ for $(z=0.5, \zeta=0.4)$
and for $(z=0.5, \zeta=0.1)$, respectively, as a function of $|\vec B|$ in
units of
GeV$^{-1}$. The value of $|\vec B|=1\;$GeV$^{-1}$ corresponds (see Fig.~1) to a
transverse distance between the two pions of the order of 0.83 fm (for
$\zeta=0.4$)
and 2.2 fm (for $\zeta=0.1$).

\vskip.1in
\noindent
{\bf 8.} In summary, we have proposed
in this note a new way to analyse data on
two pion production. For this aim, we have introduced the impact picture of
$q\,\bar q$ hadronization in a pair of mesons. We emphasize the connection of
the phase shift analysis of low energy reactions to this new approach.
More data are needed in $\gamma^*
\;\gamma$ and $\gamma^*\; p$ reactions to get a
precise transverse space picture of
exclusive hadronization processes. As in the case of the
impact picture of GPDs, the
ultimate spacial resolution is determined by the value of the virtuality
of the
photon.

Let us also emphasize possible application of our analysis to other
channels like $K^+\,K^-$ or $\rho \rho$, which can be studied along
similar lines. Final states with more particles, e.g. $\pi \pi \pi$ in
$ \gamma^* \gamma$ collisions
\cite{PT},
or the reaction
$\gamma^* N --> \gamma \pi \pi^0 N'$  \cite{Bl} will require
more study.

\vskip.3in
\hspace{-.6cm}{\Large \bf Acknowledgments}

\vskip.1in
\noindent
We acknowlege useful discussions with I.~Anikin and M.~Diehl. L.Sz.
is
grateful for the
warm hospitality at Ecole Polytechnique.
This work has been supported in part by
the french-polish collaboration agreement Polonium.

\end{document}